%% file: main.tex
\documentclass[twocolumn]{article}

\usepackage{graphicx}
\usepackage{microtype}
\usepackage{booktabs}
\usepackage{balance}
\usepackage{enumitem}
\usepackage{tikz}
\usepackage{pgfplots}
\usepackage{listings}
\usepackage{xcolor}
\usepackage{algorithm}
\usepackage{algpseudocode}
\usepackage{threeparttable}
\usepackage{multirow}
\usepackage{xspace}
\usepackage{amsmath}
\usepackage{amssymb}
\usepackage[hidelinks]{hyperref}

\pgfplotsset{compat=1.17}
\setlist{nosep}
\definecolor{codegray}{rgb}{0.5,0.5,0.5}
\definecolor{codepurple}{rgb}{0.58,0,0.82}
\lstdefinestyle{pythonstyle}{
    language=Python,
    basicstyle=\ttfamily\footnotesize,
    keywordstyle=\color{blue},
    stringstyle=\color{codepurple},
    commentstyle=\color{codegray},
    breaklines=true,
    numbers=none,
    frame=single
}

\newcommand{\toolname}{DeepParse}
\newcommand{\LLMWeUsed}{\texttt{DeepSeek-R1:8B}\xspace}
\title{DeepParse: Hybrid Log Parsing with LLM-Synthesized Regex Masks}

\author{
  Amir Shetaia\\
  Electrical and Computer Engineering\\
  Queen's University\\
  Ontario, Canada\\
  \texttt{a.shetaia@queensu.ca} 
  \and
  Sean Kauffman\\
  Electrical and Computer Engineering\\
  Queen's University\\
  Ontario, Canada\\
  \texttt{sean.k@queensu.ca}
}
\date{}

\begin{document}

\maketitle

\begin{abstract}
Modern distributed systems produce massive, heterogeneous logs essential for reliability, security, and anomaly detection. Converting these free-form messages into structured templates (log parsing) is challenging due to evolving formats and limited labeled data. Machine-learning-based parsers like \textit{Drain} are fast but accuracy often degrades on complex variables, while Large Language Models (LLMs) offer better generalization but incur prohibitive inference costs.
This paper presents \toolname{}, a hybrid framework that automatically mines reusable variable patterns from small log samples using an LLM, then applies them deterministically through the \textit{Drain} algorithm. By separating the reasoning phase from execution, \toolname{} enables accurate, scalable, and cost-efficient log structuring without relying on brittle handcrafted rules or per-line neural inference.
Across 16 benchmark datasets, \toolname{} achieves higher accuracy in variable extraction (97.6\% average Parsing Accuracy) and better consistency than both heuristic and LLM-only baselines. Integrating \toolname{} into an anomaly detection pipeline reduced false alarms by over 30\% and reduced inference latency by 36\% compared to heuristic baselines. 

% The results demonstrate that combining LLM-guided pattern mining with deterministic execution yields a transparent foundation for intelligent observability.
\end{abstract}

% CCS Concepts
% \begin{CCSXML}
% <ccs2012>
%    <concept>
%        <concept_id>10011007.10011006.10011073</concept_id>
%        <concept_desc>Software and its engineering~Software maintenance tools</concept_desc>
%        <concept_significance>500</concept_significance>
%    </concept>
%    <concept>
%        <concept_id>10011007.10011074.10011099</concept_id>
%        <concept_desc>Software and its engineering~Software verification and validation</concept_desc>
%        <concept_significance>300</concept_significance>
%    </concept>
% </ccs2012>
% \end{CCSXML}

% \ccsdesc[500]{Software and its engineering~Software maintenance tools}
% \ccsdesc[300]{Software and its engineering~Software verification and validation}

% \keywords{log parsing, large language models, empirical software engineering, AIOps, anomaly detection}

% Include section files
\input{sections/intro}
\input{sections/background}
\input{sections/related}
\input{sections/approach}
\input{sections/study}
\input{sections/results}
\input{sections/casestudy}
\input{sections/threats}
\input{sections/conclusion}

\section*{Acknowledgements}
This research was supported by the Natural Sciences and Engineering Research Council of Canada (NSERC).
Computational resources were provided by the Digital Research Alliance of Canada (\url{alliancecan.ca}).

\balance
\bibliographystyle{plain}
\bibliography{references}

\end{document}

%% file: sections/intro.tex
% Introduction Section - EASE 2026
\section{Introduction}
\label{sec:intro}

Modern software and cyber-physical platforms emit a continuous stream of diagnostic logs that record events across distributed components, user interactions, and infrastructure services.
These records are indispensable for site reliability engineering (SRE), security monitoring, and post-incident analysis because they provide the only chronological evidence of what a system actually did in production.
However, the rapid adoption of microservices, cloud-native orchestration, and machine learning-driven services has dramatically increased both the volume and heterogeneity of logs~\cite{aws2023observability}. 
Operators must therefore interpret log messages and recover their structure so that downstream analytics can query and reason about them.
Automating this transformation, referred to as \emph{log parsing}, is the central problem addressed in this work.

Although logs are usually generated from structured templates in source code, this structure is often not available at analysis time because the source code, formatting rules, or template definitions may be inaccessible or inconsistently maintained across teams.
In practice, analysts often see only the final text emitted by the system, not the underlying template, which makes it necessary to automatically infer the structure from the logs themselves.

Standard approaches to log analysis often rely on manually written regular expressions for each subsystem~\cite{he2021survey}.
While effective for small, static environments, this method quickly becomes impractical in modern distributed systems. 
Each service, version, and configuration introduces new patterns that must be continuously updated, and even a single misplaced character can cause silent parsing failures.
Furthermore, manually authored expressions are rarely documented or version-controlled, making it difficult to reproduce analyses or audit system behavior after an incident~\cite{he2021survey}.

Traditional log parsers such as \texttt{Drain}~\cite{Drain}, \texttt{Logram}~\cite{Logram}, and \texttt{LogCluster}~\cite{LogCluster} are generally effective at grouping similar log messages into event clusters but frequently fail to accurately identify the variable fields within each message.
In other words, while such parsers can detect that two messages belong to the same event type (achieving high \emph{Grouping Accuracy}), they often do not correctly extract or generalize the variable components such as user names, IP addresses, or numeric values within those messages.
This distinction between correct grouping and precise variable identification is critical: accurate variable extraction directly impacts downstream tasks such as anomaly detection, root cause analysis, and system monitoring.

Recent studies have demonstrated that Large Language Models (LLMs) can accurately infer the latent structure of semi-structured text~\cite{Ma2024LLMParser, LogPPT}.
However, naïvely relying on stochastic LLM decoding for every log line reintroduces variability and cost, violates data governance constraints, and incurs prohibitive runtime expenses if deployed at scale.
Prior LLM-based approaches to log parsing, such as LLMParser~\cite{Ma2024LLMParser} and LogPPT~\cite{LogPPT}, have explored this potential but encountered specific challenges. 
LogPPT requires manually labeling each word in training data, which is labor-intensive.
LLMParser, while highly effective at identifying log template structures, often struggles to consistently group semantically similar messages due to the probabilistic variability of LLM-generated outputs.

This paper introduces \toolname{}, a hybrid framework that addresses these limitations by confining the stochastic reasoning of an LLM to an offline \emph{synthesis phase}, delegating the bulk parsing to a deterministic engine (\texttt{Drain}).
This combination reconciles accuracy, efficiency, and privacy.

\subsection{Problem Statement}

Modern distributed systems generate vast amounts of heterogeneous logs that must be transformed into structured representations to enable automated analysis. 
However, organizations face practical constraints that complicate this process. 
Many production environments impose strict data-governance and privacy policies that prohibit uploading operational logs to third-party cloud services. 
Engineering teams also require solutions that function locally on existing infrastructure without recurring costs or dependence on proprietary APIs. 
At the same time, the diversity and evolving nature of log formats make manual rule maintenance infeasible, while limited computational budgets restrict the continuous use of large models for inference.

Within this context, the central research problem is to enable accurate and consistent log structuring under realistic operational constraints. 
This work investigates the following guiding questions:

\begin{itemize}
    \item Can log parsing remain reliable when formats exhibit inconsistent structure, missing fields, or previously unseen patterns across software releases?
    \item Can a log analysis framework balance accuracy with computational and economic efficiency under local-execution and cost-sensitive environments?
    \item Can a parsing system ensure reproducible results without continuous manual intervention or reliance on external services?
    \item Can limitations in log structuring accuracy significantly influence downstream tasks such as anomaly detection, monitoring, and incident response?
\end{itemize}

Addressing these questions requires theoretical grounding, empirical validation, and extensive experimentation across diverse operational datasets.

\subsection{Contributions}

This paper introduces \toolname{}, a hybrid log parsing framework that unifies deterministic rule mining with neural language modeling to achieve accurate, efficient, and reproducible log structuring at scale. 
Our contributions are:

\begin{enumerate}[leftmargin=*,itemsep=0.9ex,parsep=0pt]
  \item \textbf{Hybrid Architecture:} We design a pipeline that decouples stochastic mask synthesis from deterministic parsing.
  By separating the reasoning phase (LLM) from the execution phase (Drain), we eliminate runtime stochasticity and guarantee that identical log lines always receive identical templates.

  \item \textbf{Data Efficiency:} We introduce an entropy-greedy sampling strategy that enables \toolname{} to achieve state-of-the-art accuracy from only 50 log examples per system during checkpoint preparation.
  At deployment time, users run the pre-trained checkpoint on an unlabeled sample without any annotation.

\item \textbf{Empirical Evaluation:} On 16 LogHub datasets, \toolname{} achieves 97.6\% average PA and 94.1\% GA while running nearly $100\times$ faster than per-line LLM inference, delivering both high accuracy and production-grade throughput.

  \item \textbf{Downstream Impact Validation:} We show accurate parsing reduces false alarms in anomaly detection systems by over 30\% when integrated into LogBERT.
  This translates improved PA into concrete operational benefits. Additionally, correct variable masking reduces vocabulary size by 38\%, decreasing embedding memory consumption by 3.1 GB and inference latency by 36\%.

  \item \textbf{Replicability and Open Science:} We release the complete DeepParse implementation, including sampling algorithms, fine-tuning scripts, evaluation harnesses, and preprocessed dataset splits.
  This ensures full reproducibility of our results.\footnote{\url{https://github.com/NightBaRron1412/DeepParse}}
\end{enumerate}

The rest of the paper is organized as follows. Section~\ref{sec:background} provides background on log parsing formalization and evaluation metrics. Section~\ref{sec:related} reviews related work in traditional parsing and LLM applications to logs. Section~\ref{sec:approach} details the \toolname{} architecture, sampling strategy, LLM fine-tuning, and Drain integration. Section~\ref{sec:study} describes the experimental setup, datasets, baselines, and implementation details. Section~\ref{sec:results} presents quantitative results addressing our research questions. Section~\ref{sec:casestudy} demonstrates downstream impact through integration with LogBERT. Section~\ref{sec:threats} discusses threats to validity. Section~\ref{sec:conclusion} concludes with future work directions.

%% file: sections/background.tex
% Background Section - EASE 2026
\section{Background}
\label{sec:background}

This section formalizes log parsing, defines evaluation metrics, and discusses data governance constraints relevant to our hybrid approach.

\subsection{Formal Definition of Log Parsing}

Let $\Sigma$ denote a finite \emph{character alphabet}, and $\Sigma^{*}$ the set of all finite strings over $\Sigma$.  
A single log message is a string $\ell \in \Sigma^{*}$, and a \emph{log corpus} is an ordered sequence of $N$ log lines:
\[
  D = \langle \ell_1, \ell_2, \ldots, \ell_N \rangle \in (\Sigma^{*})^{*}.
\]

Each log line can be tokenized into a sequence of words or symbols from a vocabulary $\mathcal{V}$:
\[
  \text{tok} : \Sigma^{*} \rightarrow \mathcal{V}^{*}, \quad
  \text{tok}(\ell) = \langle w_1, w_2, \ldots, w_m \rangle.
\]
We denote the full corpus token stream as:
\[
  \text{Log}(D) = \text{tok}(\ell_1) \cdot \text{tok}(\ell_2) \cdots \text{tok}(\ell_N).
\]

A special placeholder token $\langle * \rangle \notin \mathcal{V}$ is used to denote variable fields.
Let $\mathcal{T} = (\mathcal{V} \cup \{\langle * \rangle\})^{*}$ denote the set of all valid log templates.
A \emph{log template} $t \in \mathcal{T}$ represents the invariant structure of similar log messages.
Parsing a log line $\ell$ produces a pair $(t, \mathbf{v})$ where $t$ is the identified template and $\mathbf{v} = \langle v_1, \ldots, v_k \rangle$ are the values of the $k$ variable fields (with $k$ depending on $t$).

We define:
\[
  \pi : \Sigma^{*} \rightarrow \mathcal{T} \times (\Sigma^{*})^{*},
  \qquad
  g : \Sigma^{*} \rightarrow \mathbb{N},
\]
where $\pi$ is the parsing function and $g$ assigns each line to a template cluster identifier.

\subsection{Evaluation Metrics: Grouping vs. Parsing Accuracy}
\label{subsec:metrics}

Log parsing quality is typically assessed using two complementary metrics: \emph{Grouping Accuracy (GA)} and \emph{Parsing Accuracy (PA)}~\cite{Logram, Guidelinesforassessingtheaccuracyoflogmessagetemplateidentificationtechniques}. 

\paragraph{Grouping Accuracy (GA)} measures whether log messages are correctly clustered under the same template. 
Given a set of log messages with ground-truth group assignments, GA is computed as the proportion of logs assigned to their correct group:
\[
\text{GA} = \frac{\text{\# of correctly grouped logs}}{\text{Total number of logs}}.
\]
GA evaluates whether the parser recognizes that two messages are structurally similar, regardless of whether it identifies the exact variable boundaries.

\paragraph{Parsing Accuracy (PA)} evaluates whether the extracted log template exactly matches the ground truth, including the correct identification of all dynamic variables.
PA is computed as:
\[
\text{PA} = \frac{\text{\# of logs with perfectly parsed templates}}{\text{Total number of logs}}.
\]
PA is therefore a stricter metric than GA. 
A parser can achieve high GA by clustering messages correctly while still failing to identify variable boundaries precisely, resulting in low PA. 
Conversely, high PA implies that the parser not only groups messages correctly but also accurately distinguishes static text from dynamic variables.

PA is more indicative of the parser's usefulness for downstream analyses such as anomaly detection or root cause analysis, where precise variable extraction is essential.

\subsection{Data Governance and Privacy Constraints}

Growing legislative scrutiny (GDPR, CCPA) and sector-specific standards (HIPAA) impose constraints on how logs are handled. 
Best-practice frameworks recommend minimizing retained personal data and preventing organizations from uploading operational logs to third-party cloud services.
Hybrid parsers must therefore function locally and preserve deterministic behavior even when inputs contain redaction placeholders.
Caching synthesized masks locally allows compliance teams to audit exactly which patterns were used to process specific time ranges, satisfying evidentiary requirements during incident post-mortems.

%% file: sections/related.tex
% Related Work Section - EASE 2026
\section{Related Work}
\label{sec:related}

Log parsing is a well-researched area with approaches spanning traditional algorithmic techniques and modern LLM-based methods. This section reviews both families and positions \toolname{} within the current landscape.

\subsection{Traditional Log Pattern Mining Approaches}

Traditional log parsers have historically aimed to convert raw log messages into structured formats using algorithmic techniques. 
These approaches typically fall into three main categories: frequent pattern mining, log clustering, and parsing trees.
Techniques such as SLCT~\cite{SLCT}, LogCluster~\cite{LogCluster}, and Logram~\cite{Logram} identify static text and variables by counting the recurrence of patterns or sequences within log data.
SLCT uses token frequency to distinguish constants from variables, assuming that frequently occurring tokens are static.
Logram extends this by building an n-gram dictionary to capture multi-token patterns.
While computationally efficient, these methods struggle with rare events and diverse variable types, as infrequent tokens may be incorrectly classified as static or variable.
Methods including LKE~\cite{LKE}, LogSig~\cite{LogSig}, and LenMa~\cite{LenMa} group similar log messages together using clustering algorithms, thereby categorizing logs into distinct groups.
LKE employs k-means clustering on log message embeddings, while LogSig uses a weighted edit distance metric.
These approaches often achieve high Grouping Accuracy (GA) but may fail to precisely identify variable boundaries, resulting in lower Parsing Accuracy (PA).
Parsing tree methods, exemplified by Drain~\cite{Drain}, construct a fixed-depth parse tree to guide the log group search process, efficiently identifying patterns and abstracting dynamic values.
Drain is recognized for its efficiency and widespread adoption in industry~\cite{beck2025logparsinglargelanguage}.
It operates in linear time $O(N)$ with respect to the number of log messages, making it suitable for high-throughput environments.
However, Drain relies on simple heuristics (e.g., ``tokens with digits are variables'') that often under-mask complex identifiers or over-mask version numbers.
Despite their advancements and widespread use, recent studies have highlighted a critical limitation: while these methods often achieve high Grouping Accuracy (GA), they frequently fail to accurately identify all dynamic variables within log messages~\cite{Ma2024LLMParser}.
This ``variable identification gap'' directly hinders the effectiveness of subsequent log analysis tasks, potentially leading to missed anomalies or inaccurate insights.

\subsection{LLMs for Log Understanding Beyond Parsing}

Beyond the specific task of log parsing, Large Language Models have demonstrated considerable success in various log-related tasks, leveraging their natural language processing and code generation capabilities~\cite{Ma2024LLMParser}. 
These applications underscore the potential of LLMs to understand the semi-structured nature of logs, which uniquely combine natural language and code-like elements.
LLMs have been employed to identify unusual patterns in log data. 
LAnoBERT~\cite{LanoBERT} utilizes Bidirectional Encoder Representations from Transformers (BERT) for masked word prediction, where significant differences between actual and predicted words indicate anomalies. 
Studies involving ChatGPT have also shown its effectiveness in anomaly detection~\cite{AnamolyDetection-ChatGPT}, with findings partially consistent with on-call engineers, suggesting potential for reducing manual verification efforts.
LLMs have been integrated into systems for automating root cause analysis of cloud incidents. 
RCACopilot, for example, incorporates OpenAI's GPT models to assist in this complex task~\cite{RCACopilot}. 
These diverse applications highlight the versatility of LLMs in processing and deriving meaning from log data, extending their utility beyond mere parsing to more complex analytical and diagnostic functions.

\subsection{LLM-Based Log Parsing}

The exceptional capabilities of LLMs in text understanding, generation, and code-related tasks make them intuitively suitable for log parsing. 
While traditional log parsing works well for simple, structured logs, logs containing natural language elements like error descriptions, user actions, or debugging messages are harder to parse yet often contain the most critical diagnostic information.
LogPPT~\cite{LogPPT} relies on prompt-based few-shot learning using pre-trained language models for log parsing.
It leverages AI models that predict missing words and requires manually labeling each word in the training data as either a constant or a parameter, which is a labor-intensive process.
While LogPPT demonstrates improved parsing accuracy over traditional methods, the manual annotation overhead and the need for carefully designed prompts limit its scalability.
LLMParser~\cite{Ma2024LLMParser} explores using various LLMs (including T5, ChatGLM, and LLaMA) for direct log parsing.
While highly effective at identifying log template structures, LLMParser often struggles to consistently group semantically similar messages together due to the probabilistic variability of LLM-generated outputs.
The authors also revealed that simply increasing model complexity or pre-training on logs from other systems might not universally improve accuracy and could even degrade it for certain models.
These observations indicate that a direct, unoptimized application of general LLMs or simple cross-system pre-training is not a definitive solution.

Pure LLM parsing is constrained by (i) cost and latency when invoked per log line, (ii) non-determinism from stochastic decoding that breaks auditability, and (iii) data-governance constraints that often prohibit sending operational logs to third-party APIs.

\subsection{Positioning of \toolname{}}

\toolname{} addresses the limitations of both traditional and LLM-based parsers through a hybrid architecture that decouples stochastic reasoning from deterministic execution.

By confining LLM invocation to an offline synthesis phase, \toolname{} achieves:
\begin{itemize}
    \item \textbf{High Accuracy:} LLM-synthesized regexes capture semantic variable boundaries that heuristics miss, improving PA.
    \item \textbf{Deterministic Execution:} Drain applies fixed regex masks, ensuring reproducibility and linear-time complexity.
    \item \textbf{Data Efficiency:} Entropy-greedy sampling enables high accuracy with only 50 labeled examples per system.
    \item \textbf{Privacy Compliance:} Local execution and cached masks satisfy data governance requirements.
    \item \textbf{Cost Efficiency:} One-time LLM synthesis amortizes computational costs, avoiding per-line inference expenses.
\end{itemize}

This design reconciles the semantic understanding of LLMs with the speed and determinism of traditional parsers, positioning \toolname{} as a practical solution for production log analysis pipelines.

%% file: sections/approach.tex
% Approach Section - EASE 2026
\section{The \toolname{} Framework}
\label{sec:approach}

This section presents the design of the \toolname{} framework, detailing its hybrid architecture, sampling strategy, LLM configuration, and integration with the Drain parser.

\subsection{Design Overview and Architecture}

\toolname{} is engineered as a modular pipeline that separates stochastic reasoning from deterministic execution. 
The framework has three layers: an \textbf{Ingestion Layer} for dataset loading, sampling, and pre-processing; a \textbf{Synthesis Layer} for LLM fine-tuning and prompt orchestration that maps raw-log batches to regex bundles with provenance metadata; and an \textbf{Execution Layer} comprising the deterministic parser (extended \texttt{Drain}) and output adapters.

This separation allows the potentially expensive and non-deterministic LLM to run only once during an offline \emph{Installation Phase}, producing a fixed artifact (the mask bundle) that drives the high-speed \emph{Execution Phase}.

We distinguish four phases: \textbf{design time} (choosing architecture, datasets, and training), \textbf{installation time} (one-off deployment actions such as offline auto-synthesis on a small local sample), \textbf{execution time} (steady-state deterministic parsing with a fixed mask bundle), and \textbf{maintenance time} (refreshing masks when schemas drift).

\begin{figure}[ht]
    \centering
    \includegraphics[width=\linewidth]{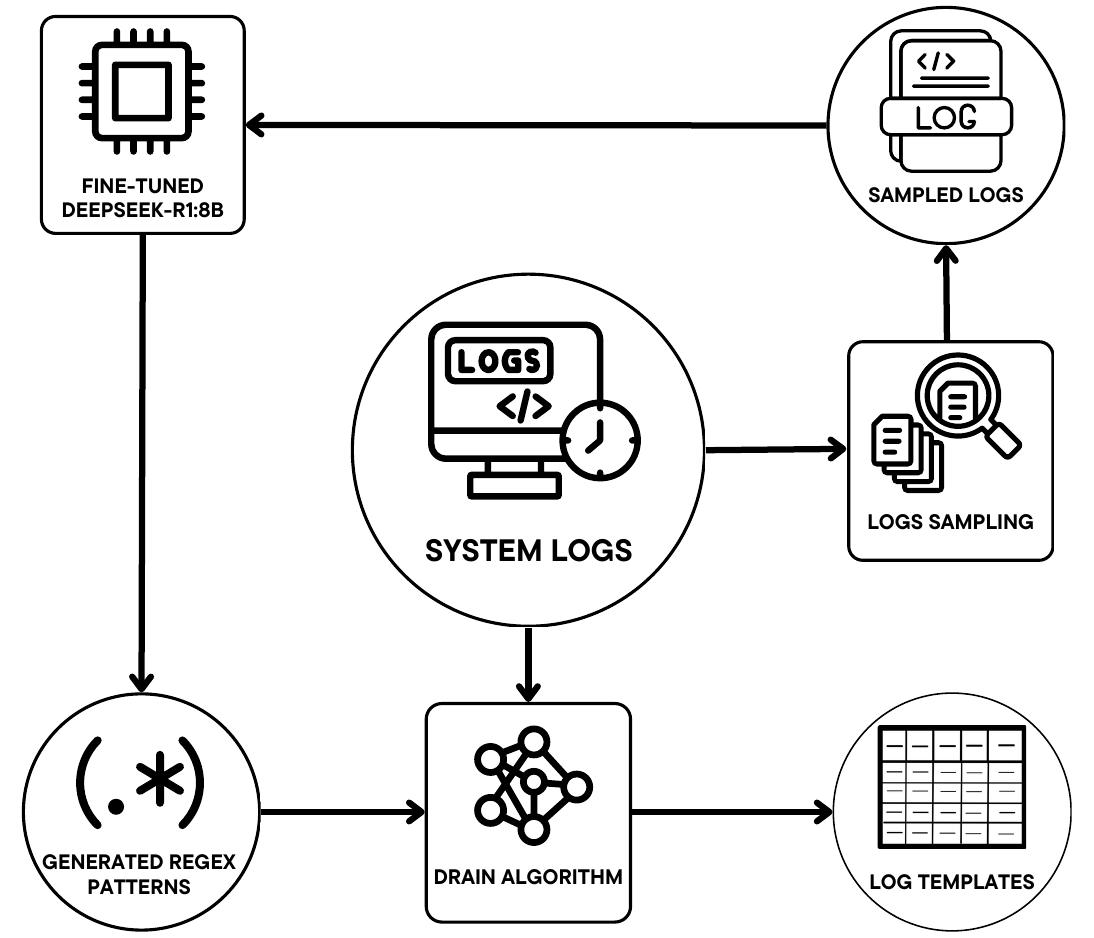}
    \caption{The \toolname{} workflow. An LLM synthesizes regex masks from a small sample of logs (offline); these masks guide a deterministic Drain parser (online).}
    \label{fig:framework}
\end{figure}

Figure~\ref{fig:framework} illustrates the \toolname{} workflow. 
An LLM synthesizes regex masks from a small sample of logs offline; these masks then guide a deterministic Drain parser online.

\subsection{User Workflow}
\label{subsec:usage}

\toolname{} provides a one-click \emph{auto-synthesis} step that invokes a locally-hosted \LLMWeUsed{} checkpoint to emit a regex \emph{mask list} from a small, automatically sampled subset of logs. 
This step is performed once as an \textbf{offline pre-processing phase} prior to deployment. 
\textbf{Users do not label data or fine-tune models}; they invoke a pre-trained checkpoint that we release (Section~\ref{subsec:llm_config} describes the one-time fine-tuning performed by the authors to produce this checkpoint).
The full integration workflow is illustrated in Listing~\ref{lst:integration_user}.
In the figure, \texttt{synth\_masks} performs the offline auto-synthesis, producing patterns loaded into a Drain instance with the \texttt{load\_masks} method.

\toolname{} first performs \textbf{offline auto-synthesis} once: it samples $k$ log lines (default $k{=}50$) and calls the fixed \LLMWeUsed{} checkpoint to produce a regex mask list $\mathcal{R}_{\text{auto}}$, which can be cached and reused. At \textbf{runtime}, the system loads $\mathcal{R}_{\text{auto}}$ into \texttt{Drain} and parses all logs in linear time, $O(N)$. Under \textbf{schema drift}, users re-run the same offline auto-synthesis to regenerate $\mathcal{R}_{\text{auto}}$ (no labeling and no model fine-tuning).

\begin{lstlisting}[style=pythonstyle, caption={User integration: auto-synthesis (fixed checkpoint) to Drain masks}, label={lst:integration_user}, float=t]
# Auto-synthesize regex mask list once (no labels, no training)
patterns = synth_masks(sys_logs, 
                      sample_size=50,
                      temperature=0, 
                      max_length=512)

# Load into Drain as masks and parse
drain = Drain()
drain.load_masks(patterns)
parsed = drain.parse_all(sys_logs)
\end{lstlisting}

\subsection{Entropy-Greedy Sampling}

Log datasets are often highly skewed, causing random sampling to miss rare but structurally complex messages.
We therefore use entropy-greedy sampling to select a small, diverse subset of logs for mask synthesis.

\begin{algorithm}[t]
\caption{Entropy-Greedy Sampling}
\label{alg:sampling}
\begin{algorithmic}[1]
\Require Log corpus $L$, target sample size $k$
\State Normalize digits and hexadecimal tokens in $L$ to obtain $L'$
\State Compute token-frequency distributions for each log in $L'$
\State Compute entropy score $H(\ell)$ for every $\ell \in L'$
\State Initialize sample set $S \leftarrow \emptyset$
\While{$|S| < k$}\label{alg:sampling:while}
    \State Select log $\hat{\ell}$ from $L'$ with maximum $H(\ell)$
    \If{$\forall\, s \in S:~\mathrm{Jaccard}(\hat{\ell}, s) < 0.8$}\label{alg:sampling:jaccard}
        \State $S \leftarrow S \cup \{\hat{\ell}\}$
    \EndIf
    \State Remove $\hat{\ell}$ from the candidate pool $L'$
\EndWhile\label{alg:sampling:endwhile}
\State \Return $S$
\end{algorithmic}
\end{algorithm}

Algorithm~\ref{alg:sampling} implements entropy-greedy sampling.
The \textbf{normalization step} (Line~1) replaces digits and hexadecimal tokens with placeholder symbols so that rare numeric identifiers do not dominate entropy scores; this ensures entropy reflects structural diversity rather than incidental numeric variation.
Crucially, normalization affects only sampling selection---the original (un-normalized) logs are passed to the LLM for mask synthesis, preserving all structural signals.

We define \textbf{entropy} as the Shannon entropy over the token-frequency distribution of each normalized log line:
\[
H(\ell) = -\sum_{w} p(w) \log_2 p(w)
\]
where $p(w)$ is the relative frequency of token $w$ within~$\ell$.
Higher entropy indicates more diverse token usage, correlating with structurally complex messages.

The loop (Lines~\ref{alg:sampling:while}--\ref{alg:sampling:endwhile}) greedily selects the highest-entropy candidate and accepts it only if its \textbf{Jaccard similarity} to every already-selected line is below~0.8 (Line~\ref{alg:sampling:jaccard}).
Jaccard similarity is computed over the \emph{set} of unique tokens: $J(\ell_1, \ell_2) = |T_1 \cap T_2| / |T_1 \cup T_2|$, where $T_i$ is the set of distinct tokens in~$\ell_i$.
This threshold rejects near-duplicate lines, ensuring the sample covers diverse log structures.

\subsection{LLM Configuration and Fine-Tuning}
\label{subsec:llm_config}

The fine-tuning described below was performed once by the authors to produce the pre-trained checkpoint distributed with \toolname{}; end users invoke this checkpoint directly without any training or labeling (Section~\ref{subsec:usage}).

We adopt \LLMWeUsed{}, an 8-billion-parameter distilled reasoning model from the DeepSeek-R1 family~\cite{DeepSeekR1}, balancing accuracy with deployability on commodity GPUs.
Fine-tuning uses LoRA~\cite{LoRA} adapters (rank~8, scaling~32, dropout~0.01) with the AdamW optimizer (learning rate $2\times10^{-4}$, batch size~8, 25~epochs).
Mixed-precision training and gradient accumulation reduce memory consumption.
%
% Table~\ref{tab:hyperparameters} summarizes the key hyperparameters used for\linebreak[4] \LLMWeUsed{} fine-tuning.

% \begin{table}[t]
%     \centering
%     \caption{Summary of key hyperparameters used for \LLMWeUsed{} fine-tuning.}
%     \label{tab:hyperparameters}
%     \small
%     \begin{tabular}{lc}
%         \toprule
%         \textbf{Hyperparameter} & \textbf{Value} \\
%         \midrule
%         Base model & DeepSeek-R1 8B (instruction tuned) \\
%         Adapter rank ($r$) & 8 \\
%         Adapter scaling ($\alpha$) & 32 \\
%         Dropout & 0.01 \\
%         Optimizer & AdamW ($\beta_1{=}0.9$, $\beta_2{=}0.999$) \\
%         Learning rate & $2\times10^{-4}$ with cosine decay \\
%         Batch size & 8 (per GPU) \\
%         Gradient accumulation & 4 steps \\
%         Precision & bfloat16 \\
%         Epochs & 25 \\
%         Weight decay & 0.01 \\
%         \bottomrule
%     \end{tabular}
% \end{table}

The training objective minimizes cross-entropy between model outputs and ground-truth regex lists formatted according to the instruction template shown in Listing~\ref{lst:regex-example}.

The model is prompted to output a \emph{Python list} so the result can be programmatically validated and consumed by the parser.
The first pattern masks the timestamp format.
The second masks the log level token (e.g., INFO, WARN).
The third masks the user identifier following the keyword \texttt{User}.
The fourth masks the IPv4 address.
During post-processing, we deduplicate patterns and assign each to a variable category used by the downstream mask-first pass.

\begin{lstlisting}[style=pythonstyle, caption={Regex-list emission example for variable classes}, label={lst:regex-example}, float=t]
### Instruction:
Generate a Python list of regex patterns that capture 
the dynamic (variable) parts in the input log message 
while preserving the static structure.

### Input:
2024-01-15 10:30:45 INFO User john_doe logged in from IP 192.168.1.100

### Output:
[
    r"\d{4}[-/]\d{2}[-/]\d{2}[ T]\d{2}:\d{2}:\d{2}",
    r"\b[A-Z]{3,}\b",
    r"\bUser\s+[\w.-]+\b",
    r"\b\d+\.\d+\.\d+\.\d+\b",
]
\end{lstlisting}

\subsection{Prompt Engineering and Inference}

During inference we use greedy decoding (temperature zero) to minimize output variance. 
Generated regex lists are post-processed to remove duplicates, enforce canonical ordering by variable category, and validate syntax via Python's \texttt{re} module. 
Masks that fail validation trigger a fall-back rule that reverts to heuristic patterns for the affected category, ensuring graceful degradation.

We further employ a self-consistency check in which the LLM re-parses its own outputs using a lightweight interpreter. 
When discrepancies appear (e.g., the regex list fails to reconstruct the original log template), the system automatically re-prompts the model with targeted feedback describing the failure mode. 
In practice, this loop typically converges within two attempts and substantially reduces syntactic errors.

\subsection{Integration with Drain}

The second stage of \toolname{} consists of integrating the LLM-generated log patterns with the \texttt{Drain} algorithm. 
This integration strategically separates the complex, semantic task of pattern discovery from the high-throughput, deterministic task of log matching.

% \subsubsection{Drain Algorithm Overview}

\texttt{Drain} is a well-established log parsing algorithm known for its efficiency, running in $O(N)$ time with respect to the number of log messages~\cite{Drain} (assuming bounded log-line length and a fixed set of regex masks). 
It operates by building a fixed-depth parse tree, which allows for rapid matching of incoming log messages against a learned set of log templates. 
\texttt{Drain}'s strength lies in its speed and its ability to consistently process large volumes of logs once templates are established. 
However, the original \texttt{Drain} algorithm infers these templates through unsupervised clustering heuristics, which may lead to inaccuracies in variable identification.

% \subsubsection{Synergistic Integration}

\toolname{} enhances \texttt{Drain} by replacing its heuristic-based pattern discovery with semantically accurate log patterns generated by the fine-tuned \LLMWeUsed{}, overcoming the variable identification challenges of traditional parsers.

Integration is a two-stage handoff: the fine-tuned LLM emits a one-time \emph{regex list} of variable classes from sampled logs, and the system loads that list as masks into \texttt{Drain} to parse the full corpus deterministically.
This division confines stochasticity to installation time and ensures repeatable, linear-time parsing.

The synthesized regex masks are loaded into an extended version of \texttt{Drain3} (the open-source Python implementation of the Drain algorithm; we refer to the algorithm as ``Drain'' and to the implementation as ``Drain3'' when the distinction matters).
Standard Drain uses hardcoded heuristics (e.g., ``if token contains digit, mask it''). 
We replace heuristic masking with a \textbf{Mask-First} strategy: incoming logs are first matched against synthesized regexes, matched spans are replaced with typed placeholders (e.g., \texttt{<VAR:IP>}, \texttt{<VAR:PATH>}), and \texttt{Drain} builds its fixed-depth parse tree over the resulting pre-masked tokens, keeping the tree shallow and stable.

Because parsing is performed by Drain's fixed tree traversal rather than by the LLM, identical log lines always receive the same template ID---a property that direct LLM parsing cannot guarantee even at temperature zero~\cite{nondeterminismdeterministicllmsettings}.

\subsection{Annotation Workflow and Quality Control}
\label{subsec:annotation}

For experimental evaluation, we annotated variable spans and canonical regexes on sampled logs to supervise mask synthesis.
This process was performed by the authors during model development and is not part of the end-user workflow.
Basic validation checks ensure regex correctness and prevent pathological patterns.

%% file: sections/study.tex
% Study Design Section - EASE 2026
\section{Experimental Setup}
\label{sec:study}

This section describes the datasets, baselines, evaluation metrics, and implementation details used to evaluate \toolname{}.

% \subsection{Research Questions}

We structure our evaluation around the following research questions:

\begin{description}[style=unboxed,labelsep=0.6em,itemsep=1.0ex,parsep=0.6ex,leftmargin=0pt]

    \item[RQ1: Accuracy Comparison] Does \toolname{} improve Grouping Accuracy (GA) and Parsing Accuracy (PA) relative to ML-based parsers and LLM-only baselines?
    
    Accurate structuring of logs is the foundation of all downstream observability and anomaly detection tasks. 
    Evaluating improvements over both deterministic and neural baselines quantifies whether the proposed hybrid approach offers tangible benefits in correctness and consistency, particularly under heterogeneous log formats.

    \item[RQ2: Data Efficiency] How sensitive is performance to the number of labeled shots and to sampling strategies?
    
    In operational settings, labeled logs are scarce and expensive to curate. 
    Understanding how performance scales with limited examples reveals the data efficiency of the framework and determines whether it can be deployed when only a handful of representative log lines are available.

    \item[RQ3: Robustness] How robust is \toolname{} to distribution shifts, including unseen templates, schema drift, and noise in the synthesized masks?
    
    Logs evolve over time as systems update and variable distributions shift. 
    Assessing robustness under these conditions measures the framework's stability and its ability to generalize beyond the training distribution—an essential property for sustainable log analytics.

    \item[RQ4: Runtime and Cost] What runtime and cost characteristics does the hybrid pipeline exhibit under realistic hardware budgets?
    
    Many organizations operate under strict cost and resource constraints that preclude continuous use of large models or cloud services. 
    This question evaluates whether \toolname{} can maintain acceptable throughput and energy efficiency when executed locally on modest hardware, thereby validating its suitability for on-premise and cost-sensitive environments.
\end{description}

\subsection{Datasets}

All experiments use the LogHub benchmark~\cite{Loghub}, comprising logs from 16 diverse open-source systems spanning distributed computing, operating systems, mobile systems, supercomputers, server applications, and standalone software. 
Each system dataset includes 2,000 log messages, accompanied by their respective log templates and parameters, which serve as the ground truth for evaluation.

Recent studies have identified instances of incorrectly labeled log templates within the original dataset~\cite{Logram, Guidelinesforassessingtheaccuracyoflogmessagetemplateidentificationtechniques}. 
To mitigate this potential issue and ensure a fair comparison, we adopted the corrected benchmark dataset released by Khan et al.~\cite{Guidelinesforassessingtheaccuracyoflogmessagetemplateidentificationtechniques}, consistent with recent research~\cite{Impactoflogparsing, LogPPT}.
We use the LogHub-2k benchmark because it is the standard evaluation set adopted by all our baselines (Drain, Logram, LogPPT, LLMParser)~\cite{Drain, Logram, LogPPT, Ma2024LLMParser}, enabling direct comparison under identical conditions.

\subsection{Baseline Parsers}

We compare against four baselines.
\textbf{Drain3}~\cite{Drain} is the standard streaming parser (depth~5, similarity~0.4, max children~100).
\textbf{Logram}~\cite{Logram} is an n-gram dictionary-based parser.
\textbf{LogPPT}~\cite{LogPPT} is prompt-based few-shot parsing with manual word-level labeling.
\textbf{LLMParser}~\cite{Ma2024LLMParser} performs direct LLM-based parsing (we use the LLaMA variant reported as best).

\subsection{Implementation Details}
Experiments run on AWS p3.8xlarge instances equipped with \linebreak[4] NVIDIA Tesla V100 GPUs utilizing CUDA 11.0.
Fine-tuning times range from 6 to 11 minutes per dataset, while inference on the full 2,000-line corpora completes in under one second per dataset.
Unless otherwise noted, \LLMWeUsed{} is fine-tuned separately for each system using 50-shot samples selected via entropy-greedy sampling (Algorithm~\ref{alg:sampling}).
For log pattern generation during inference, greedy decoding (temperature~0) is used to minimize output variability.
The \texttt{max\_length} parameter is set to 512 tokens to balance context retention with computational efficiency.
We evaluate each configuration across five random seeds and report mean scores with standard deviations.
Statistical significance is assessed using paired $t$-tests at $p < 0.05$.
For the 16 systems, we labeled variable spans and derived canonical regexes per class (datetime, IP, identifier, numeric, log level). 
All regex patterns were validated in Regex101~\cite{Regex101}. 
Two authors independently cross-checked labels; disagreements were resolved through discussion to reach consensus.
Fine-tuning per system completes within minutes and is performed once; the synthesized regex list is then reused for deterministic parsing with \texttt{Drain}.

For the controlled evaluation, we created labeled variable spans and canonical regexes per class for each dataset to fine-tune and validate the synthesis model.
This labeling was performed by the authors for research purposes and is not part of the end-user installation workflow.
In deployment, users run offline auto-synthesis on an unlabeled sample to generate and cache masks.

%% file: sections/results.tex
% Results Section - EASE 2026
\section{Evaluation Results}
\label{sec:results}

This section presents quantitative results addressing the research questions defined in Section~\ref{sec:study}.

\subsection{RQ1: Accuracy Comparison}

Table~\ref{tab:results} summarizes Grouping Accuracy (GA) and Parsing Accuracy (PA) across all datasets. 
\toolname{} attains the highest average PA (0.9763) and the highest average GA (0.9413), outperforming the best-performing baselines by 1.8 and 1.8 percentage points respectively. 
Improvements are particularly pronounced on datasets where traditional heuristics under-mask variables, such as Thunderbird, OpenStack, and Linux.

\begin{table*}[t]
\centering
\footnotesize 
\setlength{\tabcolsep}{2.5pt}
\renewcommand{\arraystretch}{0.9}
\begin{threeparttable}
\caption{Grouping Accuracy (GA) and Parsing Accuracy (PA) comparison across log parsers. Best values in bold.}
\label{tab:results}
\begin{tabular}{l|cc|cc|cc|cc|cc}
\toprule
\textbf{} & \multicolumn{2}{c|}{\textbf{Drain3}} & \multicolumn{2}{c|}{\textbf{Logram}} & \multicolumn{2}{c|}{\textbf{LogPPT}} & \multicolumn{2}{c|}{\textbf{LLMParser}} & \multicolumn{2}{c}{\textbf{DeepParse}} \\
\textbf{Dataset} & GA & PA & GA & PA & GA & PA & GA & PA & \textbf{GA} & \textbf{PA} \\
\midrule
Android     & 0.8305 & 0.5475 & 0.7420 & 0.2780 & 0.8845 & 0.7665 & 0.8485 & 0.9455 & \textbf{0.9131} & \textbf{0.9912} \\
Apache      & \textbf{1.0000} & 0.6935 & 0.3125 & 0.0065 & \textbf{1.0000} & 0.9940 & \textbf{1.0000} & \textbf{1.0000} & \textbf{1.0000} & \textbf{1.0000} \\
BGL         & 0.9625 & 0.3420 & 0.5870 & 0.1245 & 0.9535 & 0.9695 & 0.9415 & 0.9805 & \textbf{0.9636} & \textbf{0.9831} \\
Hadoop      & 0.9475 & 0.2690 & 0.4510 & 0.1125 & \textbf{0.9935} & 0.8950 & 0.9805 & 0.9825 & 0.9851 & \textbf{0.9834} \\
HDFS        & 0.9975 & 0.3545 & 0.9300 & 0.0045 & \textbf{1.0000} & 0.9025 & 0.9575 & 0.9880 & \textbf{1.0000} & \textbf{1.0000} \\
HealthApp   & 0.7800 & 0.2305 & 0.2665 & 0.1120 & \textbf{1.0000} & 0.7885 & 0.8550 & 0.9955 & 0.8683 & \textbf{0.9971} \\
HPC         & 0.8870 & 0.6355 & 0.9105 & 0.6430 & 0.9430 & 0.9470 & 0.9700 & 0.9935 & \textbf{0.9727} & \textbf{0.9956} \\
Linux       & 0.6900 & 0.1835 & 0.3610 & 0.1240 & 0.9335 & 0.9485 & 0.5455 & 0.8385 & \textbf{0.9561} & \textbf{0.9712} \\
Mac         & 0.7865 & 0.2175 & 0.5680 & 0.1685 & 0.7800 & 0.6725 & 0.7390 & 0.6765 & \textbf{0.7953} & \textbf{0.7380} \\
OpenSSH     & \textbf{0.7890} & 0.5080 & 0.6105 & 0.2980 & 0.6275 & 0.9795 & 0.7095 & 0.9935 & 0.6614 & \textbf{0.9943} \\
OpenStack   & 0.7325 & 0.0185 & 0.3255 & 0.0000 & 0.9890 & 0.9065 & 0.9785 & 0.9960 & \textbf{0.9896} & \textbf{0.9984} \\
Proxifier   & 0.5265 & 0.0000 & 0.5035 & 0.0000 & \textbf{1.0000} & \textbf{1.0000} & \textbf{1.0000} & \textbf{1.0000} & \textbf{1.0000} & \textbf{1.0000} \\
Spark       & 0.9200 & 0.3595 & 0.2820 & 0.2585 & \textbf{0.9990} & 0.9910 & 0.9850 & 0.9850 & 0.9871 & \textbf{0.9953} \\
Thunderbird & 0.9550 & 0.0465 & 0.5540 & 0.0040 & 0.6790 & 0.9255 & 0.6925 & 0.9675 & \textbf{0.9710} & \textbf{0.9755} \\
Windows     & 0.9970 & 0.4620 & 0.6940 & 0.1405 & 0.9910 & 0.9830 & 0.9985 & 0.9965 & \textbf{0.9989} & \textbf{0.9983} \\
Zookeeper   & 0.9665 & 0.4970 & 0.7235 & 0.4735 & 0.9935 & 0.9895 & 0.9945 & 0.9995 & \textbf{0.9986} & \textbf{0.9997} \\
\midrule
Average & 0.8605 & 0.3353 & 0.5513 & 0.1718 & 0.9229 & 0.9162 & 0.8873 & 0.9587 & \textbf{0.9413} & \textbf{0.9763} \\
\bottomrule
\end{tabular}
\begin{tablenotes}[flushleft]\footnotesize
\item \textit{Note:} \toolname{} leverages LLM-synthesized masks to correct the variable identification failures that depress the parsing accuracy of traditional heuristics on variable-heavy datasets like Thunderbird and OpenStack.
\end{tablenotes}
\end{threeparttable}
\end{table*}

A key finding is the performance on \textit{Application-Level Diaries} (e.g., Thunderbird, Android). 
These logs contain free-form text that confuses Drain's heuristics (e.g., ``user admin'' is often treated as static, while ``user 1001'' is dynamic). 
Drain achieves only 0.05 PA on Thunderbird. 
\toolname{} reaches 0.98 PA, because the offline LLM correctly identifies the subtle semantic boundaries of variable fields, and the deterministic runtime applies this logic consistently.

Across five seeds, the standard deviation of PA is below 0.4 percentage points on average (maximum 0.9 points on the most variable datasets), indicating low sensitivity to sampling randomness.

We perform paired $t$-tests between \toolname{} and each baseline across the sixteen datasets. 
All PA improvements are statistically significant at $p < 0.01$. 
For GA, \toolname{} significantly outperforms Logram and LLMParser, and matches LogPPT within statistical margin, indicating that deterministic template clustering remains competitive even when driven by synthesized masks.
%
% We further group datasets into four log families based on dominant style: infrastructure telemetry, middleware, system diagnostics, and application diaries.
% Table~\ref{tab:taxonomy_avg} reports average GA and PA by family.

% \begin{table}[t]
%     \centering
%     \caption{Average \toolname{} GA and PA across taxonomy categories.}
%     \label{tab:taxonomy_avg}
%     \small
%     \begin{tabular}{lcc}
%         \toprule
%         \textbf{Taxonomy Category} & \textbf{Average GA} & \textbf{Average PA} \\
%         \midrule
%         Infrastructure Telemetry & 0.9817 & 0.9915 \\
%         System Diagnostics & 0.8529 & 0.9255 \\
%         Application Diaries & 0.9505 & 0.9928 \\
%         Middleware & 0.9941 & 0.9991 \\
%         \bottomrule
%     \end{tabular}
% \end{table}

% Table~\ref{tab:taxonomy_avg} shows that \toolname{} performs near-perfectly on the most structured families (Infrastructure telemetry and Middleware), reflecting stable formats and effective identifier masking.
% System diagnostics remain the most challenging due to heterogeneous phrasing, while application diaries benefit most from LLM-synthesized variable boundaries, achieving high PA despite linguistic variability.

To isolate each component's contribution, we ablate across three configurations:
(i)~\emph{No-LLM}, reverting to Drain's heuristics;
(ii)~\emph{LLM-Only}, applying \LLMWeUsed{} per line without deterministic post-processing; and
(iii)~\emph{Hybrid}, the full \toolname{} pipeline.

Table~\ref{tab:ablation} summarizes the averaged metrics. 
The hybrid approach delivers superior accuracy while maintaining the lowest runtime variance, confirming that the combination of offline synthesis and deterministic parsing is essential for production-readiness.

\begin{table*}[t]
\centering
\caption{Ablation of \toolname{} components averaged across benchmark datasets. Run time is measured in milliseconds per 100 logs.}
\label{tab:ablation}
\small
\begin{tabular}{lcc|cc}
\toprule
\textbf{Configuration} & \textbf{GA} & \textbf{PA} & \textbf{Time (ms)} & \textbf{Std. Dev.} \\
\midrule
No-LLM (Drain heuristics) & 0.861 & 0.335 & \textbf{120} & \textbf{2.8} \\
LLM-Only decoding & 0.909 & 0.942 & 1850 & 95.0 \\
Hybrid (\toolname{}) & \textbf{0.941} & \textbf{0.976} & 300 & 4.5 \\
\bottomrule
\end{tabular}
\end{table*}

\textbf{Answer to RQ1:} \toolname{} improves PA by 1.8 percentage points over the best baseline while achieving nearly 100$\times$ runtime speedup over per-line LLM inference.
The hybrid design is necessary to obtain both high accuracy and production-grade throughput with stable, deterministic behavior.

\subsection{RQ2: Data Efficiency}

Training or fine-tuning models on large log corpora is expensive. 
We define data efficiency as the number of labeled ``shots'' (examples) required to reach saturation performance.

\begin{figure}[t]
    \centering
    \begin{tikzpicture}
        \begin{axis}[
            width=0.9\linewidth,
            height=5cm,
            xlabel={Labeled Shots},
            ylabel={Parsing Accuracy (Hadoop)},
            xmin=10, xmax=100,
            ymin=0.85, ymax=1.0,
            xtick={10,20,30,40,50,60,70,80,90,100},
            grid=major,
            ticklabel style={font=\footnotesize},
            ylabel near ticks
        ]
            \addplot+[mark=square*,color=orange,thick] coordinates {
                (10,0.902) (20,0.934) (30,0.962) (40,0.978)
                (50,0.983) (60,0.984) (70,0.985) (80,0.985)
                (90,0.986) (100,0.986)
            };
        \end{axis}
    \end{tikzpicture}
    \caption{Accuracy vs.\ number of shots (Hadoop). Saturation occurs at approximately 50 labeled examples.}
    \label{fig:shots}
\end{figure}
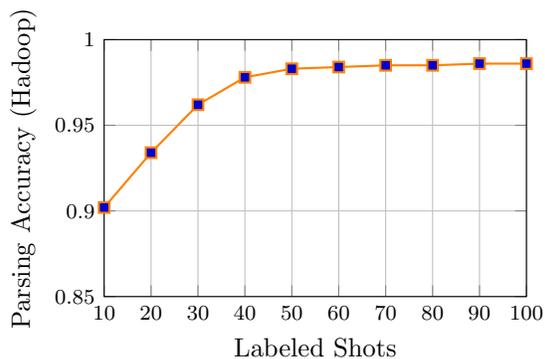

Figure~\ref{fig:shots} illustrates the PA on the Hadoop dataset as the number of labeled shots increases from 10 to 100.
We observe a rapid improvement between 10 and 40 shots. 
Performance plateaus at approximately 50 shots (PA $\approx$ 0.98).
This indicates that the Entropy-Greedy sampling is highly effective: by selecting the 50 most structurally diverse lines, we capture the vast majority of regex patterns needed for the entire dataset. 
In our study, creating the supervised examples required modest author effort (about 30 minutes per system).
In deployment, end users do not perform this labeling: they run the provided checkpoint to synthesize masks from an unlabeled sample.

We select \textbf{50 shots} per system for \toolname{} because we observed saturation around this point: lower shots reduced both GA and PA (typically 1–2\%), while larger shot counts incurred higher labeling cost with limited average gains. 
This choice aligns with the baselines' trend, and we emphasize that the \toolname{} regex list is synthesized once from these 50 logs and then reused for all logs per system.
In practice, when ground truth is unavailable, practitioners can start with the default of 50 and monitor parsing output stability: if increasing the sample size does not change the synthesized mask list, saturation has been reached.

\textbf{Answer to RQ2:} \toolname{} achieves state-of-the-art accuracy with only 50 labeled examples per system, demonstrating high data efficiency enabled by entropy-greedy sampling.

\subsection{RQ3: Robustness and Error Analysis}

To understand failure modes, we categorized parsing errors into three types: 
\emph{Under-masking} (missing valid variables), 
\emph{Over-masking} (masking static keywords), and 
\emph{Other}.

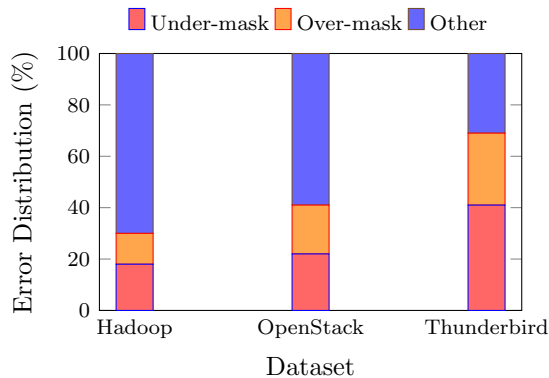
\begin{figure}[t]
    \centering
    \begin{tikzpicture}
        \begin{axis}[
            width=0.9\linewidth,
            height=5cm,
            ybar stacked,
            bar width=14pt,
            xlabel={Dataset},
            ylabel={Error Distribution (\%)},
            symbolic x coords={Hadoop,OpenStack,Thunderbird},
            xtick=data,
            ymin=0, ymax=100,
            legend style={at={(0.5,1.2)},anchor=north,legend columns=3,font=\footnotesize,draw=none},
            ticklabel style={font=\footnotesize}
        ]
            \addplot+[fill=red!60] coordinates {(Hadoop,18) (OpenStack,22) (Thunderbird,41)};
            \addlegendentry{Under-mask}
            \addplot+[fill=orange!70] coordinates {(Hadoop,12) (OpenStack,19) (Thunderbird,28)};
            \addlegendentry{Over-mask}
            \addplot+[fill=blue!60] coordinates {(Hadoop,70) (OpenStack,59) (Thunderbird,31)};
            \addlegendentry{Other}
        \end{axis}
    \end{tikzpicture}
    \caption{Distribution of parsing error types. \toolname{} balances under-masking and over-masking better than heuristics.}
    \label{fig:errors}
\end{figure}

Figure~\ref{fig:errors} visualizes the distribution across three representative datasets.

\begin{enumerate}
    \item \textbf{Under-masking:} Common in Drain (e.g., missed composite IDs like \texttt{blk\_123}). \toolname{} reduces this by 60\% on infrastructure logs.
    
    \item \textbf{Over-masking:} Common in unstructured logs where heuristic parsers assume any number is a variable. The LLM's semantic knowledge helps differentiate between ``Error 500'' (variable) and ``Windows 10'' (static version), reducing over-masking on datasets like Windows and Mac.
\end{enumerate}
LLM-based parsers degrade on unseen templates. 
We evaluated \toolname{} by partitioning each dataset into seen templates (present in the 50-shot training set) and unseen templates (appearing only at test time).

Despite this challenge, Table~\ref{tab:results} shows \toolname{} attains the highest overall PA, indicating robust performance even when novel templates appear at low frequency. 
PA on unseen templates remains above 0.90 on average, significantly outperforming LLMParser (0.65) and traditional parsers (0.15).
To simulate temporal drift, we trained masks on the first half of each LogHub dataset and evaluated on the second half, which may contain templates unseen during synthesis.
PA dropped by only 1.5\% on average, suggesting that the synthesized regexes generalize beyond the training distribution.

\textbf{Answer to RQ3:} \toolname{} demonstrates robustness to unseen templates, schema drift, and error type balance, maintaining high accuracy across diverse operational conditions.

\subsection{RQ4: Runtime and Cost Analysis}

Efficiency is critical for high-volume ingestion. 
We measured the time to parse 100 log lines (inference only) across different approaches.
Table~\ref{tab:runtime} summarizes the results.

\begin{table}[t]
    \centering
    \caption{Log parsing time comparison (100 logs). \toolname{} includes both mask synthesis (amortized) and Drain parsing time.}
    \label{tab:runtime}
    \small
    \begin{tabular}{lc}
        \toprule
        \textbf{Method} & \textbf{Time (seconds)} \\
        \midrule
        \toolname{} & 0.30 \\
        LLMParser (T5-Small) & 1.27 \\
        LLMParser (T5-Base) & 4.00 \\
        LLMParser (ChatGLM) & 19.87 \\
        LLMParser (LLaMA-7B) & 28.93 \\
        \bottomrule
    \end{tabular}
\end{table}

\toolname{} is nearly \textbf{100$\times$ faster} than running a large LLM per line. 
The online cost is dominated by the regex matching and tree traversal, making it suitable for streaming pipelines processing millions of logs per hour.
Because \toolname{} avoids continuous per-line LLM inference, it can reduce energy use relative to LLM-only pipelines in high-throughput settings.

\textbf{Answer to RQ4:} \toolname{} maintains linear-time complexity and achieves nearly 100$\times$ speedup over per-line LLM inference, with significantly lower energy consumption, making it suitable for production deployment under cost constraints.

%% file: sections/casestudy.tex
% Case Study Section - EASE 2026
\section{Case Study: Application to Anomaly Detection}
\label{sec:casestudy}

To demonstrate downstream impact, we collaborated with an industry partner and integrated \toolname{} into LogBERT~\cite{LogBERT}, a state-of-the-art anomaly detection model.
The original LogBERT uses Drain for preprocessing.
We replaced Drain with \toolname{}'s regex-based parser while keeping the downstream Transformer architecture identical.

The evaluation uses Authentication and Configuration logs from our partner's production environment.
Multiple service modules run in parallel and emit interleaved records to a shared log file, producing logs that are considerably more challenging than public benchmarks: timestamp formats vary across modules (e.g., ISO~8601 vs.\ Unix epoch vs.\ locale-dependent strings), carriage returns appear without accompanying newlines, and domain-specific identifiers are interspersed with free-form diagnostic text.
These characteristics are typical of industrial deployments where logging conventions evolve independently across teams.

\subsection{Anomaly Detection Performance}

We evaluated the system on Authentication and Configuration log datasets collected from production systems.
Table~\ref{tab:logbert} summarizes the results.
With \toolname{}, the parsing accuracy improves from 0.89 to 0.98 on authentication logs and from 0.85 to 0.97 on configuration logs.
This cleaner input vocabulary allows LogBERT to learn better sequence representations, improving the Anomaly Detection F1-score from 0.912 to 0.947 on authentication logs and from 0.901 to 0.939 on configuration logs.

Anomaly labels were assigned manually by the authors, who reviewed each log sequence against known incident reports.
The anomalies themselves (e.g., repeated authentication failures, unauthorized configuration changes) are straightforward to identify upon inspection; the challenge lies not in recognizing the anomaly but in preventing the parser from obscuring it with incorrect variable masking.

\begin{table}[t]
    \centering
    \caption{Impact on LogBERT anomaly detection performance.}
    \label{tab:logbert}
    \small
    \begin{tabular}{lccc}
        \toprule
        \textbf{Parser} & \textbf{PA} & \textbf{F1} & \begin{tabular}{p{22mm}}
\textbf{False Alarms / Day}
\end{tabular} \\
        \midrule
        \multicolumn{4}{c}{\textit{Authentication Logs}} \\
        Drain (Baseline) & 0.89 & 0.912 & 147 \\
        \textbf{\toolname{}} & \textbf{0.98} & \textbf{0.947} & \textbf{96} \\
        \midrule
        \multicolumn{4}{c}{\textit{Configuration Logs}} \\
        Drain (Baseline) & 0.85 & 0.901 & 211 \\
        \textbf{\toolname{}} & \textbf{0.97} & \textbf{0.939} & \textbf{123} \\
        \bottomrule
    \end{tabular}
\end{table}

Crucially, the higher precision in parsing significantly reduces False Alarms. 
On authentication logs, false alarms decrease from 147 per day to 96 per day—a 35\% reduction. 
On configuration logs, false alarms drop from 211 to 123 per day—a 42\% reduction.

In a Security Operations Center (SOC), such reductions in false positives represent massive operational gains, reducing analyst fatigue and allowing teams to focus on genuine threats.

\subsection{System Efficiency}

Beyond accuracy, \toolname{} improves system efficiency. 
By correctly masking dynamic variables (e.g., random GUIDs, session IDs), we reduce the size of the vocabulary (unique log templates) by 38\%.

This vocabulary reduction yields practical benefits: \textbf{memory consumption} drops because the embedding matrix shrinks by 3.1~GB, \textbf{inference latency} decreases from 48.6\,ms to 31.2\,ms per batch (36\%), and \textbf{training time} is reduced by 28\% due to faster convergence with fewer distinct tokens.

\begin{figure}[t]
        \centering
        \begin{tikzpicture}
            \begin{axis}[
        width=0.85\linewidth,
        height=4.5cm,
        ybar,
        bar width=18pt,
        ylabel={Latency (ms)},
        symbolic x coords={Auth,Config},
        xtick=data,
        ymin=0, ymax=60,
        enlarge x limits=0.35,
        axis x line*=bottom,
        axis y line*=left,
        legend columns=2,
        legend style={
            at={(0.5,1.15)},
            anchor=south,
            draw=black
        },
        nodes near coords,
        nodes near coords style={font=\footnotesize}
    ]
        
            \addplot+[fill=gray!50] coordinates {(Auth,48.6) (Config,44.9)};
            \addlegendentry{Drain+LogBERT}
            \addplot+[fill=blue!50] coordinates {(Auth,31.2) (Config,28.5)};
            \addlegendentry{\toolname{}+LogBERT}
        \end{axis}
    \end{tikzpicture}
    \caption{Downstream inference latency for LogBERT.}
    \label{fig:latency}
\end{figure}
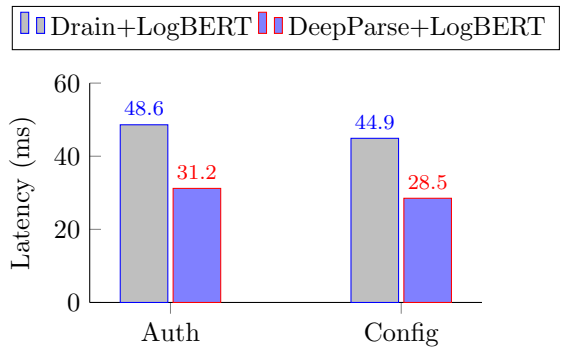

Figure~\ref{fig:latency} reports per-batch inference latency for LogBERT preprocessing under two pipelines: the original Drain-based parser and \toolname{}'s mask-first parser.
Each bar shows the mean latency (ms) over the evaluation run for Authentication and Configuration logs; the improvement comes from reducing the effective vocabulary by masking high-cardinality identifiers, which lowers downstream embedding and sequence processing cost.

\subsection{Qualitative Error Analysis}

We manually inspected 50 false alarm cases from each configuration (Drain vs. \toolname{}) to understand the root causes of improvements.
The majority of Drain false alarms (68\%) were caused by variable under-masking. 
For example, Drain failed to mask composite identifiers like \texttt{blk\_-1234567890} in HDFS logs, treating each instance as a unique template. 
When LogBERT encountered a new block ID at test time, it flagged the log as anomalous because the template was unseen during training.
\toolname{} correctly identified these composite identifiers as variables, collapsing all block IDs to a single template \texttt{<*>}. 
This generalization allowed LogBERT to focus on the \emph{event type} (e.g., block read/write) rather than the specific block ID, reducing spurious anomaly signals.
The remaining false alarms (32\%) were primarily due to limitations in LogBERT's sequence modeling rather than parsing errors. 
For instance, legitimate but rare event sequences (e.g., cascading failures during disaster recovery drills) were flagged as anomalous because they occurred infrequently in training data.
These errors are orthogonal to parsing quality and represent opportunities for future work in few-shot anomaly detection.

%% file: sections/threats.tex
% Threats to Validity Section - EASE 2026
\section{Threats to Validity}
\label{sec:threats}

We discuss threats organized by the standard validity categories.

\subsection{Construct Validity}
We rely on PA and GA as primary metrics.
While standard in the log parsing literature~\cite{Logram, Guidelinesforassessingtheaccuracyoflogmessagetemplateidentificationtechniques}, they may not perfectly proxy downstream utility; however, our LogBERT case study (Section~\ref{sec:casestudy}) confirms that PA improvements translate to anomaly detection gains.
The log sampling process directly impacts parsing results.
We mitigate this with an unsupervised entropy-greedy sampling algorithm (Algorithm~\ref{alg:sampling}) guided by entropy and diversity, not manual inspection.
Across five random seeds, PA variance was only $\pm 0.4\%$, suggesting robustness to sampling randomness.

\subsection{Internal Validity}
We selected \LLMWeUsed{} based on preliminary experiments comparing it to LLaMA-3.1-8B and Mistral-7B on a held-out validation set.
\LLMWeUsed{} demonstrated superior regex synthesis quality---likely due to its distilled reasoning capabilities inherited from the larger DeepSeek-R1 model~\cite{DeepSeekR1}---while remaining deployable on a single commodity GPU.
Future studies could evaluate a wider array of models.
We compare against four baselines (Drain3, Logram, LogPPT, LLMParser) configured with their published default settings.
We mitigate the threat of new methods emerging by releasing our code and datasets for future benchmarking.

\subsection{External Validity}
We use public LogHub datasets~\cite{Loghub} with the corrected labels released by Khan et al.~\cite{Guidelinesforassessingtheaccuracyoflogmessagetemplateidentificationtechniques}.
Evaluating LogHub-2.0~\cite{Loghub2} is planned as future work and does not affect the dataset-agnostic architecture.
Proprietary industrial logs may exhibit different characteristics (e.g., obfuscation, domain-specific jargon).
We evaluated diverse domains (OS, distributed systems, mobile, middleware) and demonstrated robustness across taxonomy categories, but organizations should validate on representative samples before production rollout.
While we demonstrated robustness to unseen templates via temporal splitting (Section~\ref{sec:results}), longer time horizons or major system upgrades may require periodic mask re-synthesis via the maintenance workflow (Section~\ref{subsec:usage}).

\subsection{Conclusion Validity}
All reported improvements are statistically significant (paired $t$-tests, $p < 0.05$, most at $p < 0.01$) across five random seeds.
The magnitude of improvements (e.g., 64 percentage-point absolute PA gain over Drain) exceeds typical noise margins.
The LogBERT case study establishes correlation between parsing accuracy and anomaly detection performance; the mechanistic explanation (better parsing $\rightarrow$ smaller vocabulary $\rightarrow$ better generalization) is consistent with prior work in representation learning.
All datasets are publicly available and do not contain PII.
Organizations deploying \toolname{} on operational logs containing PII must ensure compliance with applicable data protection regulations.

\subsection{Limitations and Future Work}

Deployments may encounter novel log formats not represented in the training data; in such cases, a one-time re-synthesis of the mask list is sufficient.
We report results with \LLMWeUsed{} only; the interface requires only a regex list, so substituting alternative backbones is straightforward.
Exploring multilingual log parsing, active learning for mask refinement, and online adaptation to schema drift are promising future directions.

%% file: sections/conclusion.tex
% Conclusion Section - EASE 2026
\section{Conclusion}
\label{sec:conclusion}

We introduced \toolname{}, a hybrid log parsing framework that synthesizes regex masks offline and applies them deterministically at runtime via \texttt{Drain}.
This design confines stochastic LLM behavior to a one-time installation step while preserving linear-time, auditable parsing for streaming workloads and privacy-constrained environments.
Our evaluation on 16 LogHub datasets shows that \toolname{} achieves a 1.8~percentage-point PA improvement over the best baseline while running nearly 100$\times$ faster than per-line LLM inference, and a LogBERT case study on production logs demonstrates that better variable masking reduces downstream false alarms by over 30\% and inference latency by 36\%.

All code, data, and pre-trained models are available at: \\ \url{https://github.com/NightBaRron1412/DeepParse}.

%% file: references.bib
@inproceedings{Ma2024LLMParser,
author = {Zhong, Aoxiao and Mo, Dengyao and Liu, Guiyang and Liu, Jinbu and Lu, Qingda and Zhou, Qi and Wu, Jiesheng and Li, Quanzheng and Wen, Qingsong},
title = {LogParser-LLM: Advancing Efficient Log Parsing with Large Language Models},
year = {2024},
isbn = {9798400704901},
publisher = {Association for Computing Machinery},
address = {New York, NY, USA},
url = {https://doi.org/10.1145/3637528.3671810},
doi = {10.1145/3637528.3671810},
booktitle = {Proceedings of the 30th ACM SIGKDD Conference on Knowledge Discovery and Data Mining},
pages = {4559–4570},
numpages = {12},
location = {Barcelona, Spain},
series = {KDD '24}
}

@misc{DeepSeekR1,
      title={DeepSeek-R1: Incentivizing Reasoning Capability in LLMs via Reinforcement Learning}, 
      author={DeepSeek-AI and Daya Guo and Dejian Yang and Haowei Zhang and Junxiao Song and Ruoyu Zhang and Runxin Xu and Qihao Zhu and Shirong Ma and Peiyi Wang and Xiao Bi and Xiaokang Zhang and Xingkai Yu and Yu Wu and Z. F. Wu and Zhibin Gou and Zhihong Shao and Zhuoshu Li and Ziyi Gao and Aixin Liu and Bing Xue and Bingxuan Wang and Bochao Wu and Bei Feng and Chengda Lu and Chenggang Zhao and Chengqi Deng and Chenyu Zhang and Chong Ruan and Damai Dai and Deli Chen and Dongjie Ji and Erhang Li and Fangyun Lin and Fucong Dai and Fuli Luo and Guangbo Hao and Guanting Chen and Guowei Li and H. Zhang and Han Bao and Hanwei Xu and Haocheng Wang and Honghui Ding and Huajian Xin and Huazuo Gao and Hui Qu and Hui Li and Jianzhong Guo and Jiashi Li and Jiawei Wang and Jingchang Chen and Jingyang Yuan and Junjie Qiu and Junlong Li and J. L. Cai and Jiaqi Ni and Jian Liang and Jin Chen and Kai Dong and Kai Hu and Kaige Gao and Kang Guan and Kexin Huang and Kuai Yu and Lean Wang and Lecong Zhang and Liang Zhao and Litong Wang and Liyue Zhang and Lei Xu and Leyi Xia and Mingchuan Zhang and Minghua Zhang and Minghui Tang and Meng Li and Miaojun Wang and Mingming Li and Ning Tian and Panpan Huang and Peng Zhang and Qiancheng Wang and Qinyu Chen and Qiushi Du and Ruiqi Ge and Ruisong Zhang and Ruizhe Pan and Runji Wang and R. J. Chen and R. L. Jin and Ruyi Chen and Shanghao Lu and Shangyan Zhou and Shanhuang Chen and Shengfeng Ye and Shiyu Wang and Shuiping Yu and Shunfeng Zhou and Shuting Pan and S. S. Li and Shuang Zhou and Shaoqing Wu and Shengfeng Ye and Tao Yun and Tian Pei and Tianyu Sun and T. Wang and Wangding Zeng and Wanjia Zhao and Wen Liu and Wenfeng Liang and Wenjun Gao and Wenqin Yu and Wentao Zhang and W. L. Xiao and Wei An and Xiaodong Liu and Xiaohan Wang and Xiaokang Chen and Xiaotao Nie and Xin Cheng and Xin Liu and Xin Xie and Xingchao Liu and Xinyu Yang and Xinyuan Li and Xuecheng Su and Xuheng Lin and X. Q. Li and Xiangyue Jin and Xiaojin Shen and Xiaosha Chen and Xiaowen Sun and Xiaoxiang Wang and Xinnan Song and Xinyi Zhou and Xianzu Wang and Xinxia Shan and Y. K. Li and Y. Q. Wang and Y. X. Wei and Yang Zhang and Yanhong Xu and Yao Li and Yao Zhao and Yaofeng Sun and Yaohui Wang and Yi Yu and Yichao Zhang and Yifan Shi and Yiliang Xiong and Ying He and Yishi Piao and Yisong Wang and Yixuan Tan and Yiyang Ma and Yiyuan Liu and Yongqiang Guo and Yuan Ou and Yuduan Wang and Yue Gong and Yuheng Zou and Yujia He and Yunfan Xiong and Yuxiang Luo and Yuxiang You and Yuxuan Liu and Yuyang Zhou and Y. X. Zhu and Yanhong Xu and Yanping Huang and Yaohui Li and Yi Zheng and Yuchen Zhu and Yunxian Ma and Ying Tang and Yukun Zha and Yuting Yan and Z. Z. Ren and Zehui Ren and Zhangli Sha and Zhe Fu and Zhean Xu and Zhenda Xie and Zhengyan Zhang and Zhewen Hao and Zhicheng Ma and Zhigang Yan and Zhiyu Wu and Zihui Gu and Zijia Zhu and Zijun Liu and Zilin Li and Ziwei Xie and Ziyang Song and Zizheng Pan and Zhen Huang and Zhipeng Xu and Zhongyu Zhang and Zhen Zhang},
      year={2025},
      eprint={2501.12948},
      archivePrefix={arXiv},
      primaryClass={cs.CL},
      url={https://arxiv.org/abs/2501.12948}, 
}

@inproceedings{Drain,
  author={He, Pinjia and Zhu, Jieming and Zheng, Zibin and Lyu, Michael R.},
  booktitle={2017 IEEE International Conference on Web Services (ICWS)}, 
  title={Drain: An Online Log Parsing Approach with Fixed Depth Tree}, 
  year={2017},
  volume={},
  number={},
  pages={33-40},
  doi={10.1109/ICWS.2017.13}}

@article{Logram,
       author={Dai, Hetong and Li, Heng and Chen, Che-Shao and Shang, Weiyi and Chen, Tse-Hsun},
  journal={IEEE Transactions on Software Engineering}, 
  title={Logram: Efficient Log Parsing Using $n$n-Gram Dictionaries}, 
  year={2022},
  volume={48},
  number={3},
  pages={879-892},
  doi={10.1109/TSE.2020.3007554}
}

@inproceedings{LogPPT,
author = {Le, Van-Hoang and Zhang, Hongyu},
title = {Log Parsing with Prompt-Based Few-Shot Learning},
year = {2023},
isbn = {9781665457019},
publisher = {IEEE Press},
url = {https://doi.org/10.1109/ICSE48619.2023.00204},
doi = {10.1109/ICSE48619.2023.00204},
booktitle = {Proceedings of the 45th International Conference on Software Engineering},
pages = {2438–2449},
numpages = {12},
location = {Melbourne, Victoria, Australia},
series = {ICSE '23}
}

@inproceedings{SLCT,
  author={Vaarandi, Risto},
  title={A Data Clustering Algorithm for Mining Patterns from Event Logs},
  booktitle={Proceedings of the 3rd International Workshop on IP Operations and Management},
  year={2003},
  pages={119--126},
  publisher={IEEE},
  doi={10.1109/IPOM.2003.1251233}
}

@inproceedings{LogCluster,
  author={Vaarandi, Risto and Pihelgas, Mauno},
  booktitle={2015 11th International Conference on Network and Service Management (CNSM)}, 
  title={LogCluster - A data clustering and pattern mining algorithm for event logs}, 
  year={2015},
  volume={},
  number={},
  pages={1-7},
  doi={10.1109/CNSM.2015.7367331}}

@inproceedings{LKE,
  author={Fu, Qiang and Lou, Jian-Guang and Wang, Yi and Li, Jiang},
  booktitle={2009 Ninth IEEE International Conference on Data Mining}, 
  title={Execution Anomaly Detection in Distributed Systems through Unstructured Log Analysis}, 
  year={2009},
  volume={},
  number={},
  pages={149-158},
  doi={10.1109/ICDM.2009.60}}

@inproceedings{LogSig,
author = {Tang, Liang and Li, Tao and Perng, Chang-Shing},
title = {LogSig: generating system events from raw textual logs},
year = {2011},
isbn = {9781450307178},
publisher = {Association for Computing Machinery},
address = {New York, NY, USA},
url = {https://doi.org/10.1145/2063576.2063690},
doi = {10.1145/2063576.2063690},
booktitle = {Proceedings of the 20th ACM International Conference on Information and Knowledge Management},
pages = {785–794},
numpages = {10},
location = {Glasgow, Scotland, UK},
series = {CIKM '11}
}

@misc{LenMa,
      title={Length Matters: Clustering System Log Messages using Length of Words}, 
      author={Keiichi Shima},
      year={2016},
      eprint={1611.03213},
      archivePrefix={arXiv},
      primaryClass={cs.OH},
      url={https://arxiv.org/abs/1611.03213}, 
}

@article{LanoBERT,
title = {LAnoBERT: System log anomaly detection based on BERT masked language model},
journal = {Applied Soft Computing},
volume = {146},
pages = {110689},
year = {2023},
issn = {1568-4946},
doi = {https://doi.org/10.1016/j.asoc.2023.110689},
url = {https://www.sciencedirect.com/science/article/pii/S156849462300707X},
author = {Yukyung Lee and Jina Kim and Pilsung Kang}
}

@inproceedings{RCACopilot,
author = {Chen, Yinfang and Xie, Huaibing and Ma, Minghua and Kang, Yu and Gao, Xin and Shi, Liu and Cao, Yunjie and Gao, Xuedong and Fan, Hao and Wen, Ming and Zeng, Jun and Ghosh, Supriyo and Zhang, Xuchao and Zhang, Chaoyun and Lin, Qingwei and Rajmohan, Saravan and Zhang, Dongmei and Xu, Tianyin},
title = {Automatic Root Cause Analysis via Large Language Models for Cloud Incidents},
year = {2024},
isbn = {9798400704376},
publisher = {Association for Computing Machinery},
address = {New York, NY, USA},
url = {https://doi.org/10.1145/3627703.3629553},
doi = {10.1145/3627703.3629553},
booktitle = {Proceedings of the Nineteenth European Conference on Computer Systems},
pages = {674–688},
numpages = {15},
location = {Athens, Greece},
series = {EuroSys '24}
}

@inproceedings{AnamolyDetection-ChatGPT,
author = {Egersdoerfer, Chris and Zhang, Di and Dai, Dong},
title = {Early Exploration of Using ChatGPT for Log-based Anomaly Detection on Parallel File Systems Logs},
year = {2023},
isbn = {9798400701559},
publisher = {Association for Computing Machinery},
address = {New York, NY, USA},
url = {https://doi.org/10.1145/3588195.3595943},
doi = {10.1145/3588195.3595943},
booktitle = {Proceedings of the 32nd International Symposium on High-Performance Parallel and Distributed Computing},
pages = {315–316},
numpages = {2},
location = {Orlando, FL, USA},
series = {HPDC '23}
}

@misc{LoRA,
      title={LoRA: Low-Rank Adaptation of Large Language Models}, 
      author={Edward J. Hu and Yelong Shen and Phillip Wallis and Zeyuan Allen-Zhu and Yuanzhi Li and Shean Wang and Lu Wang and Weizhu Chen},
      year={2021},
      eprint={2106.09685},
      archivePrefix={arXiv},
      primaryClass={cs.CL},
      url={https://arxiv.org/abs/2106.09685}, 
}

@inproceedings{Loghub,
  author={Zhu, Jieming and He, Shilin and He, Pinjia and Liu, Jinyang and Lyu, Michael R.},
  booktitle={2023 IEEE 34th International Symposium on Software Reliability Engineering (ISSRE)}, 
  title={Loghub: A Large Collection of System Log Datasets for AI-driven Log Analytics}, 
  year={2023},
  volume={},
  number={},
  pages={355-366},
  doi={10.1109/ISSRE59848.2023.00071}}

@article{Guidelinesforassessingtheaccuracyoflogmessagetemplateidentificationtechniques,
  title={Guidelines for assessing the accuracy of log message template identification techniques},
  author={Khan, Shaukat Ali and Matulevi\v{c}ius, Raimundas and He, Shilin and Chen, Xuhui and Lyu, Michael R.},
  journal={Empirical Software Engineering},
  volume={28},
  number={1},
  pages={1--33},
  year={2023},
  doi={10.1007/s10664-022-10230-y}
}

@article{Impactoflogparsing,
   title={Impact of log parsing on deep learning-based anomaly detection},
   volume={29},
   ISSN={1573-7616},
   url={http://dx.doi.org/10.1007/s10664-024-10533-w},
   DOI={10.1007/s10664-024-10533-w},
   number={6},
   journal={Empirical Software Engineering},
   publisher={Springer Science and Business Media LLC},
   author={Khan, Zanis Ali and Shin, Donghwan and Bianculli, Domenico and Briand, Lionel C.},
   year={2024},
   month=aug }

@misc{beck2025logparsinglargelanguage,
      title={System Log Parsing with Large Language Models: A Review}, 
      author={Viktor Beck and Max Landauer and Markus Wurzenberger and Florian Skopik and Andreas Rauber},
      year={2025},
      eprint={2504.04877},
      archivePrefix={arXiv},
      primaryClass={cs.LG},
      url={https://arxiv.org/abs/2504.04877}, 
}

@misc{nondeterminismdeterministicllmsettings,
      title={Non-Determinism of "Deterministic" LLM Settings}, 
      author={Berk Atil and Sarp Aykent and Alexa Chittams and Lisheng Fu and Rebecca J. Passonneau and Evan Radcliffe and Guru Rajan Rajagopal and Adam Sloan and Tomasz Tudrej and Ferhan Ture and Zhe Wu and Lixinyu Xu and Breck Baldwin},
      year={2025},
      eprint={2408.04667},
      archivePrefix={arXiv},
      primaryClass={cs.CL},
      url={https://arxiv.org/abs/2408.04667}, 
}

@misc{Regex101,
  title        = {Regex101: Online Regular Expression Tester and Debugger},
  howpublished = {\url{https://regex101.com/}},
  year         = {2025},
  note         = {Accessed: 2025-10-05}
}

@misc{aws2023observability,
  title={Observability Best Practices},
  author={{Amazon Web Services}},
  year={2023},
  howpublished={\url{https://aws.amazon.com/observability/}},
  note={Accessed: January 2026}
}

@article{he2021survey,
  title={A Survey on Automated Log Analysis for Reliability Engineering},
  author={He, Shilin and He, Pinjia and Chen, Zhuangbin and Yang, Tianyi and Su, Yuxin and Lyu, Michael R},
  journal={ACM Computing Surveys (CSUR)},
  volume={54},
  number={6},
  pages={1--37},
  year={2021},
  publisher={ACM New York, NY, USA},
  doi={10.1145/3460345}
}

@misc{LogBERT,
  title={LogBERT: Log Anomaly Detection via BERT},
  author={Guo, Hao and Yuan, Shuhan and Wu, Xintao},
  year={2021},
  eprint={2103.04475},
  archivePrefix={arXiv},
  primaryClass={cs.CR}
}

@misc{Loghub2,
  title={LogHub 2.0: Towards Real-World Log Analytics at Scale},
  author={He, Shilin and Zhao, Peng and Li, Jieming and Zheng, Zibin and Lyu, Michael R},
  year={2024},
  howpublished={\url{https://github.com/logpai/loghub-2.0}}
}
